\documentclass[apjl]{emulateapj}

\usepackage{natbib}
\usepackage{amsmath,amssymb}
\usepackage{mathptmx}
\usepackage{graphicx}

\newcommand  \ergs     {\ifmmode {\rm erg\,s}^{-1} \else erg s$^{-1}$\fi}
\newcommand  \Msunyr     {\ifmmode {\rm \Msun\,yr}^{-1} \else \Msun\ yr$^{-1}$\fi}
\newcommand  \msunyr     {\ifmmode {\rm \Msun\,yr}^{-1} \else \Msun\ yr$^{-1}$\fi}
\newcommand  \cmii     {\ifmmode {\rm cm}^{-2} \else cm$^{-2}$\fi}
\newcommand  \cmiii     {\ifmmode {\rm cm}^{-3} \else cm$^{-3}$\fi}

\def\Hubble{\ifmmode {\rm km\,s}^{-1}\,{\rm Mpc}^{-1}\else km\,s$^{-1}$\,Mpc$^{-1}$\fi}
\def\Msun{\ifmmode M_{\odot} \else $M_{\odot}$\fi}
\def\Lsun{\ifmmode L_{\odot} \else $L_{\odot}$\fi}
\def\Zsun{\ifmmode Z_{\odot} \else $Z_{\odot}$\fi}

\def\qo{\ifmmode q_{0} \else $q_{0}$\fi}
\def\Ho{\ifmmode H_{0} \else $H_{0}$\fi}
\def\ho{\ifmmode h_{0} \else $h_{0}$\fi}
\def\qo{\ifmmode q_{0} \else $q_{0}$\fi}
\def\ao{\ifmmode a_{0} \else $a_{0}$\fi}
\def\to{\ifmmode t_{0} \else $t_{0}$\fi}
\def\omm{\ifmmode \Omega_{{\rm M}} \else $\Omega_{{\rm M}}$\fi}
\def\omlam{\ifmmode \Omega_{\Lambda} \else $\Omega_{\Lambda}$\fi}

\def\gtsim{\raisebox{-.5ex}{$\;\stackrel{>}{\sim}\;$}}

\def\mgii{\ifmmode {\rm Mg}{\textsc{ii}} \else Mg\,{\sc ii}\fi}
\newcommand \MgII {\ifmmode {\rm Mg}\,{\sc ii}\,\lambda2798 \else Mg\,{\sc ii}\,$\lambda2798$\fi}
\def\Hbeta{\ifmmode {\rm H}\beta \else H$\beta$\fi}

\newcommand{\lbol}  {\ifmmode L_{\rm bol} \else $L_{\rm bol}$\fi}
\newcommand{\lagn}  {\ifmmode L_{\rm AGN} \else $L_{\rm AGN}$\fi}
\newcommand{\lsf}   {\ifmmode L_{\rm SF} \else $L_{\rm SF}$\fi}
\newcommand{\LHD}   {\ifmmode L_{\rm HD} \else $L_{\rm HD}$\fi}
\newcommand{\lledd} {\ifmmode L/L_{\rm Edd} \else $L/L_{\rm Edd}$\fi}
\newcommand{\fwmg}  {\ifmmode {\rm FWHM}\left(\mgii\right) \else FWHM(\mgii)\fi}
\newcommand{\CFHD}  {\ifmmode {\rm CF}_{\rm HD} \else ${\rm CF}_{\rm HD}$\fi}
\newcommand{\mbh}   {\ifmmode M_{\rm BH} \else $M_{\rm BH}$\fi}
\newcommand{\mstar}   {\ifmmode M_{*} \else $M_{*}$\fi}

\def  \mic         {$\mu$m}
\def  \MgII         {\ifmmode {\rm Mg}\,{\sc ii}\,\lambda2798
                  \else Mg\,{\sc ii}\,$\lambda2798$\fi}
\def  \mgii         {\ifmmode {\rm Mg}\,{\sc ii} \else Mg\,{\sc ii}\fi}

\def \herschel {{\it Herschel}}
\def\Chisq{\ifmmode \chi^{2} \else $\chi^{2}$}
\def\z48{$z\simeq4.8$}

\begin{document}


\title{Extreme star formation in the host galaxies of the fastest growing super-massive black holes at z=4.8}

\author{Rivay Mor\altaffilmark{1},
Hagai Netzer\altaffilmark{1},
Benny Trakhtenbrot\altaffilmark{1},
Ohad Shemmer\altaffilmark{2},
\& Paulina Lira\altaffilmark{3}
}

\altaffiltext{1}
{School of Physics and Astronomy and the Wise Observatory,
The Raymond and Beverly Sackler Faculty of Exact Sciences,
Tel-Aviv University, Tel-Aviv 69978, Israel}

\altaffiltext{2}
{Department of Physics, University of North Texas, Denton, TX 76203, USA}

\altaffiltext{3}
{Departamento de Astronomia, Universidad de Chile, Camino del Observatorio 1515, Santiago, Chile}

\email{rivay@wise.tau.ac.il}


\begin{abstract}
We report new \herschel\footnote
{\herschel\ is an ESA space observatory with science instruments provided by European-led Principal Investigator consortia and with important participation from NASA.}
observations of 25 \z48 extremely luminous optically selected active galactic nuclei (AGNs).
Five of the sources have extremely large star forming (SF) luminosities, \lsf, corresponding to SF rates (SFRs) of 2800--5600 \Msunyr\ assuming a Salpeter IMF.
The remaining sources have only upper limits on their SFRs but stacking their \herschel\ images results in a mean SFR of $700\pm 150$ \msunyr.
The higher SFRs in our sample are comparable to the highest observed values so far, at any redshift.
Our sample does not contain obscured AGNs, which enables us to investigate several evolutionary scenarios connecting super-massive black holes and SF activity in the early universe. 
The most probable scenario is that we are witnessing the peak of SF activity in some sources and the beginning of the post-starburst decline in others.
We suggest that all 25 sources, which are at their peak AGN activity, are in large mergers.
AGN feedback may be responsible for diminishing the SF activity in 20 of them
but is not operating efficiently in 5 others.

\end{abstract}




\section{Introduction}
The growth and evolution of super-massive black holes (SMBHs) are closely connected to the formation and evolution of their host galaxies.
While SMBHs grow through accretion of matter from their surroundings during an active galactic nucleus (AGN) phase, their host galaxies grow by star formation (SF).
The bolometric luminosity of the AGN (\lagn) is found to be related to the SF luminosity (\lsf; defined as the integrated luminosity due to SF between 8--1000 \mic) in its host galaxy.
In AGN-dominated sources (those with $\lagn > \lsf$), there is evidence for a simple power-law relationship between the two luminosities that
can be expressed as $\lsf\simeq10^{43}(\lagn/(10^{43} \ergs))^{0.7}$ \cite[]{Netzer2009}. 
This relationship is found in type-I \cite[]{Netzer+07_QUEST2,Lutz2008} and type-II \cite[]{Netzer2009} sources. 

Recent studies provide more insight into the general \lagn--\lsf\ dependence. 
\cite{Shao2010} and \cite{Hatziminaoglou2010} measured \lsf\ for the hosts of known AGNs in the GOODS-N and HerMES fields, respectively.
Both studies find indications that in SF-dominated sources (i.e. $\lsf\gtsim\lagn$) there is no clear correlation between the two luminosities.
Furthermore, \cite{Shao2010} found that for SF-dominated sources, in a given redshift bin, \lsf\ is roughly constant with increasing \lagn.
This behavior can be interpreted as an indication for different stages of evolution, where in sources with $\lsf>\lagn$ 
the SMBH has not yet reached its highest accretion rate phase. 
These \herschel-based studies included mostly low \lagn\ sources and very few sources at $z\geq$2.5. 
At higher redshifts and higher \lagn\ there are very few known SF-dominated sources  
although a handful of sources with \lsf$\simeq$\lagn\ have been found mainly by sub-mm observations
\cite[e.g.,][]{Isaak2002, Priddey2003,Wu2009,Leipski2010}. 

Galaxy evolution scenarios suggest two modes of SF \cite[see e.g.,][]{Rodighiero2011}. 
The steadier process of secular evolution is common in isolated disk galaxies and can reach SF rates (SFRs) of $\sim 400$ \Msunyr\ at high redshift.
A less common process with SFR that can exceed $\sim$1000 \Msunyr, is associated with mergers between two large  
gas-rich galaxies \cite[][]{DiMatteo2005,Guyon2006,Sijacki2011,Valiante2011}.
Both processes result in cold gas inflow into the center of the system which can trigger AGN activity. 
Numerical simulations and semi-analytic models of mergers suggest that in such events, the fastest SMBH growth phase succeeds the 
peak of SF activity by several hundred Myr and is likely to take place when the SMBH is obscured \cite[e.g.,][]{Hopkins2006, DiMatteo2008}.
If correct, it means that the most luminous SF phase of the merger precedes the most luminous AGN phase.
Observationally this is supported by the fact that sub-mm galaxies (SMGs) have high \lsf, however, they often exhibit little or no AGN activity.

SF and AGN activity may also be related through AGN feedback. This process can diminish or even terminate SF and SMBH accretion through fast winds and 
intense AGN radiation \cite[e.g.,][]{DiMatteo2005, Springel2005, Sijacki2007}. 
The power of AGN feedback depends on \lagn\ and can be important in the final stages of both SF modes.

In this {\it Letter} we report new \herschel\ observations of 25 optically selected AGNs from our \z48, flux-limited sample \cite[][hereafter T11]{Trakhtenbrot2011}.
Our results provide evidence for extreme SFRs in 20\% of the sources, indicating merger-driven SF activity. 
In section~\ref{sec_observations} we describe the sample and the observations and explain our method for deriving \lsf\ and SFR.
Section~\ref{sec_results} compares our findings with earlier results and discusses the way they improve the 
understanding of the relations between AGN activity and SF in the
early universe. 
Throughout this {\it Letter} we assume $\Ho=70$ \Hubble, $\omm = 0.3$
and $\omlam = 0.7$. 

\begin{table*}[t]
  \caption{\lsf\ and SFRs for the five individually detected and
stacked sources}
  \label{tab:lsf}
  \begin{center}
    \begin{tabular}{cccccccc} \hline \hline
    Object name & Redshift & \lagn$^a$  & $f_{250}$ & $f_{350}$ &
$f_{500}$ &\lsf$^b$        & SFR$^c$ \\
                &          & ($10^{47} \ergs$) & mJy & mJy & mJy
&($10^{13} \Lsun$) &  (\Msun yr$^{-1}$)    \\
    \hline
J0331-0741 & 4.73 & 1.23 & 27.6 $\pm$ 6.2 & 29.1 $\pm$ 6.7 & 19.1
$\pm$ 7.0 & $2.18_{-0.27}^{+0.38}$ & $3771_{-472}^{+653}$ \\ 
J0807+1328 & 4.88 & 1.17 & 13.5 $\pm$ 5.9 & 22.7 $\pm$ 6.5 & 21.3
$\pm$ 7.0 & $1.65_{-0.25}^{+0.45}$ & $2860_{-431}^{+787}$ \\ 
J1341+0141 & 4.69 & 1.82 & 34.5 $\pm$ 6.4 & 44.0 $\pm$ 7.1 & 37.0
$\pm$ 7.4 & $3.26_{-0.28}^{+0.39}$ & $5645_{-479}^{+667}$ \\ 
J1619+1238 & 4.81 & 0.95 & 39.4 $\pm$ 6.5 & 32.9 $\pm$ 6.8 & 21.2
$\pm$ 7.0 & $3.00_{-0.44}^{+0.20}$ & $5184_{-759}^{+346}$ \\ 
J2225-0014 & 4.89 & 1.70 & 23.3 $\pm$ 6.1 & 26.2 $\pm$ 6.6 & 18.6
$\pm$ 6.9 & $2.18_{-0.39}^{+0.29}$ & $3771_{-669}^{+495}$ \\ 
\hline
Stacked source & 4.75 & 0.78 & $4.53_{-1.02}^{+1.24}$ &
$6.22_{-1.75}^{+1.62}$ & $3.65_{-1.31}^{+1.43}$ &
$0.32_{-0.25}^{+0.25}$ & $697_{-312}^{+359}$ \\ 
    \hline \hline
   \multicolumn{8}{l}{$^a$ \begin{footnotesize}Values taken from T11
except J1619+1238 for which \lagn was calculated as described in T11.
\end{footnotesize}}  \\
   \multicolumn{8}{l}{$^b$ \begin{footnotesize}Assuming that there is
no confusing source within 10$''$ from the AGN\end{footnotesize}}  \\
   \multicolumn{8}{l}{$^c$ \begin{footnotesize}Assuming a Salpeter
IMF\end{footnotesize}}  \\

    \end{tabular}
  \end{center}
\end{table*}



\section{Observations and Data Analysis}
\label{sec_observations}

\subsection{Basic Measurements}
The 25 sources presented here are part of our sample of 40 \z48
luminous, unobscured, AGNs selected from the Sloan Digital Sky Survey
(SDSS) and described in detail in T11.
The sources were selected from the SDSS/DR6 database, requiring
$4.65<z<4.92$ and $f_{\lambda}(1450$~\AA)$>6\times10^{-18}\,\ergs$
\AA$^{-1}$, to ensure sufficient H-band brightness.
Since the sources were optically selected they are not biased with regards to SF properties.
The redshift of 4.8 was chosen to allow the measurement of the \MgII\ emission line and the AGN continuum luminosity at rest-frame 3000 \AA, using H-band spectroscopy.
These allow a reliable measurement of \mbh\ and normalized accretion rate, \lledd\ \cite[]{McLure2004a}. 
Based on this sample T11 showed that the highest luminosity AGNs at \z48\ have, on average, 
lower \mbh\ and higher \lledd\ than the highest luminosity AGNs in the corresponding samples of sources at $z\simeq$2--3.5 from \cite{Shemmer2004} and \cite{Netzer2007a}. 
Most of the \z48\ SMBHs seem to be at the end of their first continuous growth phase that started at $z\gtsim$10 
and are on their way to become the most massive ($>10^{10}$ \Msun) black holes in the Universe.
An additional object, J1619+1238, was selected as part of the $z\simeq4.8$ campaign described in T11, 
but the low quality its H-band spectrum prohibited a reliable \mbh\ estimate, and thus it was not included in T11.

So far, 33 sources of the total of 41 \z48\ sources have been
observed with the SPIRE instrument \cite[]{Griffin2010} on-board
\herschel, providing images at 250, 350, and 500 \mic, corresponding
to the rest-frame far infrared (FIR) wavelength range of 43-86 \mic. 
All SPIRE observations were made in the small-map mode, most suitable for point sources. 
The data reduction process starts with the level 0.5 product of the SPIRE pipeline. 
We then apply the standard tools, which are provided by the \herschel\ Science Centre (HSC) via the HIPE
software \cite[][version 7.3]{Ott2010}, and using the version 7.0 of calibration files.
Since all of our sources appear as point sources in the images, we follow the guidelines of the HSC and apply a peak fitting method in order to measure the total flux. 
This is done by fitting a two dimensional fixed-width Gaussian function to the image and taking the value at its peak to be the flux of the source.

In seven sources, the peak emission is shifted by 10--30$''$ from the
optical location of the AGN. The low resolution SPIRE images do not
allow proper source separation in such cases and deep observations at
shorter wavelengths are needed to confirm these detections. A {\it
Spitzer} campaign is underway to secure such observations, therefore
we defer the analysis of these sources to a future publication. 
Of the remaining 26 sources, five were detected at above a 3-$\sigma$
significance level in at least two SPIRE bands, and twenty were not
detected at all SPIRE bands.
One additional source (J1306+0236) was detected above a
3-$\sigma$ level only in the 350 \mic\ band. 
In this {\it Letter} we focus on two
groups of sources either detected at least in two bands or undetected
in all three bands. Therefore, we defer the analysis of this source to
a future publication and hereafter refer only to the remaining 25
sources.

There are three sources of uncertainty related to the SPIRE images 
(see the SPIRE observers` manual\footnote{http://herschel.esac.esa.int/Docs/SPIRE/html/spire\_om.html}).
The first uncertainty is in the fitted value and includes instrument and confusion noise. 
The integration time of each observation (either 222 or 296 seconds) is long enough to minimize the instrument noise
and obtain an image that is dominated by the extragalactic confusion noise. 
The confusion noise is estimated to be 5.8, 6.3 and 6.8 mJy/beam at 250, 350 and 500 \mic, respectively \cite[]{Nguyen2010}.
The pixelization correction of the images introduces an uncertainty of about 2--3\% of the flux density. 
An additional uncertainty is associated with the calibration process and is about 7\% of the flux density.
These uncertainties are added in quadrature and listed together with the measured fluxes in table~\ref{tab:lsf}.

\subsection{SED Fitting and Stacking Analysis}
\label{sec_modeling}

\begin{figure}[t]
\includegraphics[width=9cm]{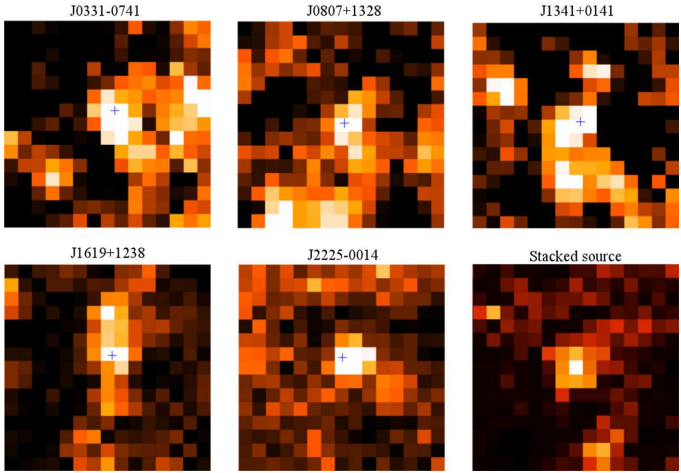}
\caption{
350 \mic\ images of the detected and stacked sources (see text). 
Each image is $2.5' \times 2.5'$ and the blue cross represents the
optical (SDSS) location of each of the individually detected sources.
}
\label{fig:images}
\end{figure}

\begin{figure}[t]
\includegraphics[width=9cm]{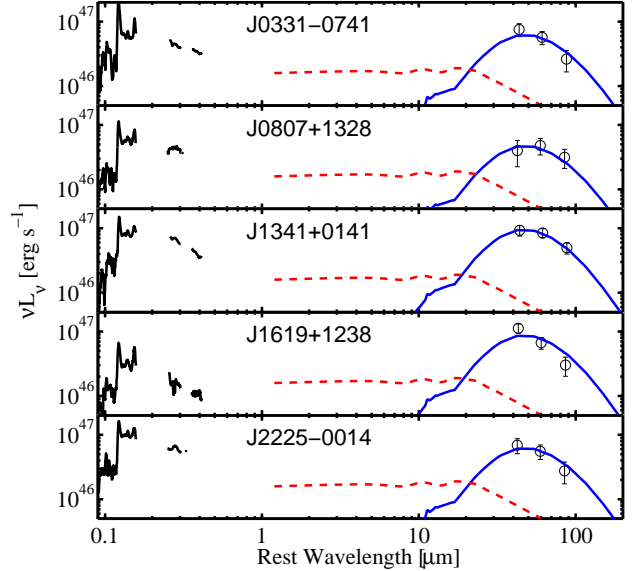}
\caption{
Spectral energy distributions (SEDs) of the 5 individually detected
sources.
The \herschel/SPIRE detections are shown as circles in each panel, and
the best fit luminous SF templates are shown as solid blue lines.
The black lines at the short wavelength end represent SDSS spectra
supplemented with H- and K-band spectroscopy (where available) from
\cite{Trakhtenbrot2011}.
Red dashed lines are mean intrinsic MIR AGN SEDs from \cite{Mor2012a}
(see text). 
}
\label{fig:SEDs}
\end{figure}

The 350 \mic\ images of the five detected sources are shown in
Figure~\ref{fig:images}, and their spectral energy distributions
(SEDs) are shown in Figure~\ref{fig:SEDs}.
We also show the SDSS, H-, and K-band (where available) spectroscopy from T11
and the mean intrinsic mid infrared (MIR) AGN SED from \cite{Mor2012a}. 
The normalization of the MIR SED uses the known \lagn\ and assumes a total covering factor by hot and warm dust of 0.5.
As discussed in \cite{Mor2012a}, this value is close to the upper limit for such luminous AGNs.
The AGN contribution to the emission at wavelengths longer than $\sim$30 \mic\ is small and its effect on the measurement of \lsf\ is negligible.

To measure \lsf\ we fit the \herschel\ data points of each of the detected sources using
a grid of templates that span a wide range of FIR luminosities and SFRs \cite[]{Chary2001}.
The FIR luminosity is the only free parameter for this set of templates.
\lsf\ is calculated by integrating the best fit model between 8 and
1000 \mic.
Comparison with other types of models which are often used to
estimate FIR luminosities (e.g., gray body) is beyond the scope of
this {\it Letter}.
The fitting procedure uses a standard \Chisq\ minimization to determine the best fit template.
The uncertainty is calculated using the standard confidence levels for \Chisq\ with two degrees of freedom.

As noted, twenty sources were not detected above 3-$\sigma$ level at
any of the bands. 
To get an average luminosity of these sources, we applied a stacking analysis to all twenty images in each band.
We first cut each image to a small stamp symmetrically around the center of the pixel in which the optical location of the source lies. 
All stamps have an equal number of pixels and are approximately 1'$\times$1' in size.
A stacked image is constructed by assigning the images with weights according to their respective exposure times, and averaging the images pixel by pixel. 
Since the dominant source of uncertainty is the extragalactic confusion noise, the slightly different exposure times have a negligible effect.
The stacking procedure revealed a statistically significant signal at all three bands (hereafter ``stacked source'') that represents the average flux of the individually undetected sources.
We then measure the flux of the stacked source by fitting a two dimensional Gaussian function as described above.

\begin{figure}[t]
\includegraphics[width=9cm]{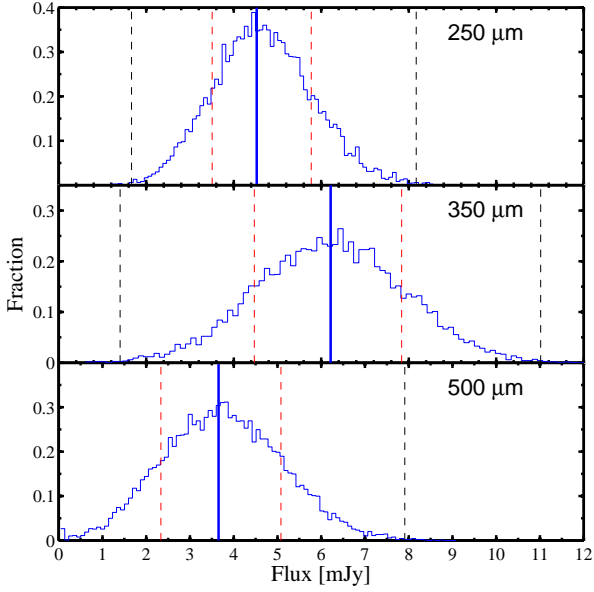}
\caption{
Probability distribution functions of the fluxes of the stacked
sources (thin solid blue lines).
The vertical solid lines mark the peaks (the most likely value) of the
distributions. 
Inner and outer dashed lines represent the different percentiles
corresponding to 1-$\sigma$ and 3-$\sigma$ uncertainties,
respectively.}
\label{fig:pdfs}
\end{figure}

The averaging of the images assumes that all objects are located at the center.
Any contribution from neighboring sources would be significantly reduced due to the fact that such contributions 
are expected to be randomly distributed in the images.
The average values can be biased if the (undetected) sources have very different fluxes. 
In such a case, few sources that are just below the confusion noise limit might skew the result towards a higher average flux. 
To overcome this we used a bootstrap approach to estimate the true value of the flux of the stacked source and its uncertainty. 
Out of the 20 images we choose 10000 random multisets of length 20. In each multiset, images may appear more than once.
We then stacked each multiset of images and measured the flux of the emergent stacked source. 
The probability distribution functions (PDFs) of the bootstrap procedure, in all three SPIRE bands, are shown in Figure~\ref{fig:pdfs}. 
The final value of the flux in each band is taken to be the maximum likelihood value of the corresponding distribution.
Its uncertainty is estimated by measuring the 16th and 84th percentiles, which are assumed to represent the 1-$\sigma$ error.
The measured fluxes of the stacked source are listed in Table~\ref{tab:lsf}.
To measure the \lsf\ of the stacked source we follow the same fitting procedure described above.


\section{Results and Discussion}
\label{sec_results}
Table~\ref{tab:lsf} lists \lagn\ (taken from T11), \lsf, and SFRs for the detected and stacked sources.
SFRs were calculated from \lsf\ assuming a Salpeter \citeyearpar{Salpeter1955} initial mass function (IMF) and using  
SFR$=1.73\times10^{-10} \lsf/\Lsun$ \cite[]{Kennicutt1998}. 
A top-heavy Kroupa IMF \cite[]{Kroupa2001} would result in SFRs that are lower by a factor of $\sim$1.6. 

We have collected from the literature FIR and sub-mm measurements of other luminous AGNs that imply $\lsf>5\times10^{12}\,\Lsun$.
Eight sources at $z\sim$4 have a single 850 \mic\ detection \cite[]{Isaak2002}.
Four sources have been reported by \cite{Priddey2003}.
Two of these at $z\sim$5 and $z\sim$6 have a single 850 \mic\ detection and 2 others, at similar redshifts,
have both 850 \mic\ and 250 GHz detections.
Eight additional sources at $z\sim$5.8--6.2 were detected at 250 GHz \cite[]{Wang2008,Wang2011}.
Finally, \cite{Leipski2010} reported multi-band measurements of two sources at $z\simeq$4.7 and 6.4.
We used the fluxes reported in these papers and applied our fitting method to obtain \lsf. 
Fitting SF templates to one or two data points is problematic in two ways.
First, a single data point fit does not allow the calculation of a meaningful confidence limit.
Second, the 850 \mic\ and 250 GHz bands translate to about 170 \mic\ rest-frame wavelength at the reported redshifts. 
This wavelength is far from the peak wavelength emission of cool dust.
To estimate the range of possible \lsf\ for the sources with a single data point, we fitted the single-band 
measurement with a gray body with emissivity index of $\beta=1.5$ and two temperatures, 40 and 60~K.
These temperatures represent the uncertainty in \lsf\ for these sources.
\lagn\ for these sources were calculated using the continuum flux density at 1450 $\AA$ and the relation 
$\log \left({\lagn}/{10^{46} \ergs}\right) = 0.94\log \left({L_{1450}}/{10^{46} \ergs}\right)+0.53$, where $L_{1450}= \lambda L_{\lambda}(1450\,\AA)$.
This relation is based on several samples of high-z, high \lagn\ sources \cite[][and T11]{Shemmer2004, Netzer2007a}.

\begin{figure}[t]
\includegraphics[width=9cm]{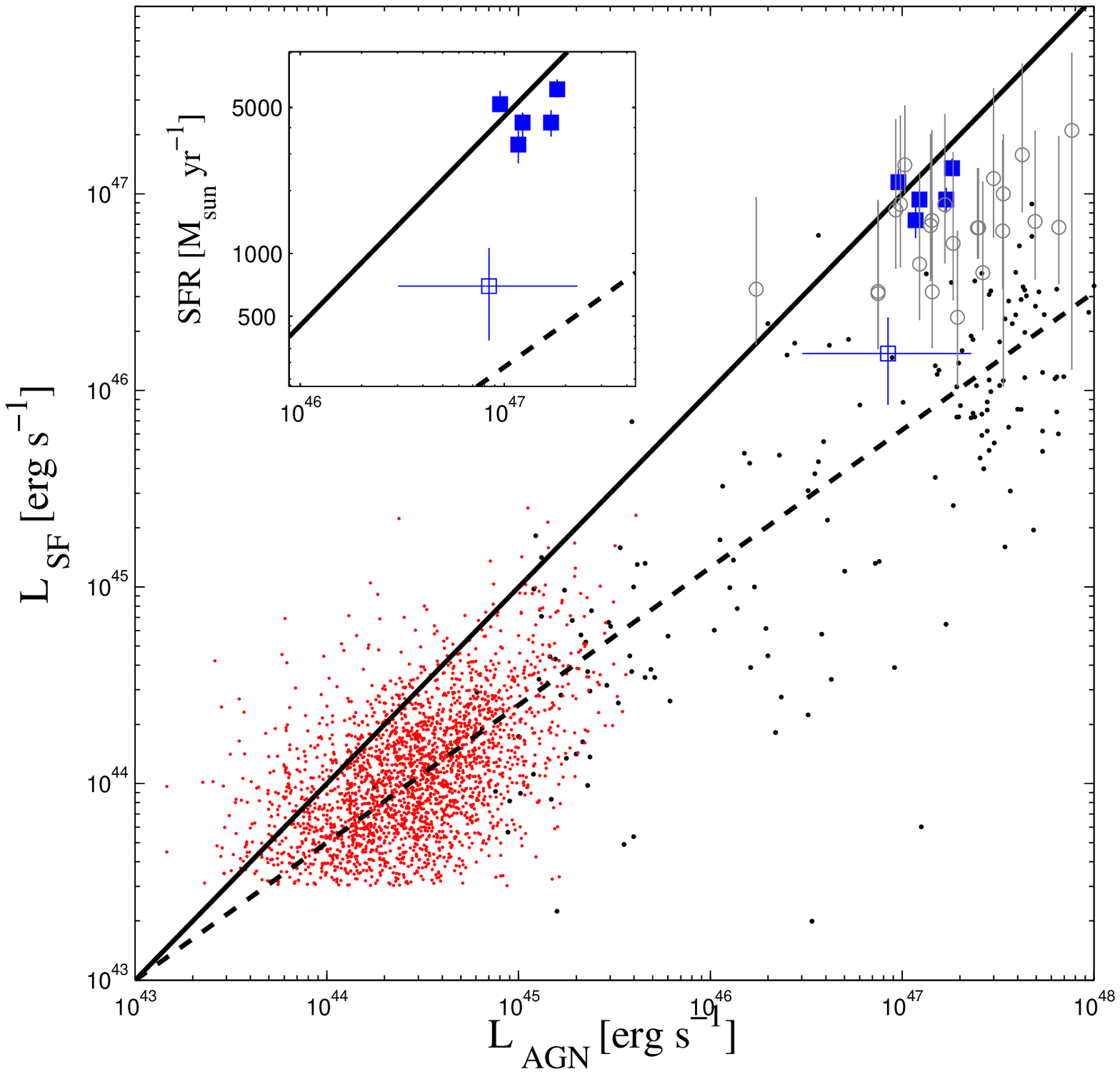}
\caption{
AGN luminosity, \lagn, vs. star formation luminosity, \lsf\ for several samples.
The five \z48\ sources detected using the \herschel/SPIRE are shown as blue squares.
The value of the stacked source is represented by the blue empty square. The horizontal error bar represents the range of \lagn\ for
the group of undetected sources, and the vertical error bar is set by the 3-$\sigma$ confidence level of the PDFs in Fig~\ref{fig:pdfs}.
Dots represent data from several samples of type-I AGN at low and high redshifts (black dots) and type-II AGN (smaller red dots), see text.
The main correlation for AGN dominated sources is shown as a dashed line and can be expressed as $\lsf \simeq 10^{43} (\lagn/(10^{43} \ergs))^{0.7}$. 
A 1:1 ratio is shown as a solid line for comparison.
Open gray circles represent observations collected from the literature. 
Their \lsf\ determination is based, in most cases, on a single sub-mm point (see text).
The insert shows \lagn\ vs. SFR for the new \z48\ data.\\
}
\label{fig:lagn_lsf}
\end{figure}

Fig.~\ref{fig:lagn_lsf} summarizes the main results of this study on the \lsf\ vs. \lagn\ diagram.
It shows a collection of several samples of low and intermediate luminosity AGN-dominated sources from \cite{Netzer2009}, 
the high luminosity high redshift sources collected from the literature, and our \z48\ sources. 
The AGN-dominated correlation line and a 1:1 line are also shown.


%
%

The two main results of our work are best explained by considering the data shown in the insert of Fig.~\ref{fig:lagn_lsf}.
First, the SFRs of the five individually detected sources are comparable to the most FIR luminous objects known, including SMGs.
Thus, the most intense SF phase in AGN hosts is not necessarily associated with an obscured SMBH phase.
Second, the detected and stacked sources occupy two distinct locations on the \lagn--\lsf\ plane.
The mean SFR of the detected sources is higher than the SFR of the stacked source by about a factor of 5.
This result is consistent with the findings of \cite{Wang2011} for a sample of z$\sim$6 quasars.
All the sources in our sample have similar optical-UV spectra, span a narrow range of \mbh\ values, similar high values of \lagn, 
and almost identical redshifts (see Table~\ref{tab:lsf}). 
Thus, the AGN properties do not explain the different SF properties.

We consider three scenarios to explain the observed different SFRs in the two groups. 
\begin{enumerate}
\item 
The different locations represent different evolutionary routes. 
The detected sources are in major mergers with extremely high SFRs and the undetected ones go through a calmer secular evolution with lower SFRs. 
The similar AGN properties and different SFRs of the sources in both groups imply different relative growth rates of the SMBH and stellar mass in each group.
Assuming that the ratios of growth rates remain the same for some time (before or after z=4.8),
this will lead to different \mstar/\mbh\ at the end of their evolution. 
Since all the sources are probably on their way to become the most massive black holes in the universe (T11), 
this scenario is not supported by observations in the local universe.

\item
All the sources are in merging systems where an initial burst of SF is followed by accretion onto one or two SMBHs. 
The different SFRs may be explained by different conditions in the mergers (e.g. galaxy size and gas content). 
If this is correct, the SFRs of the sources should have been more evenly distributed across the SFR range. 
Furthermore, in this case the gas supply to the SMBH is only loosely connected with the SF in the host. 

\item
All the sources reside in merging systems of similar type and the different SFRs are due to different stages of the merger process.
The process can continue in two different ways.
The first is that both SFR and \lledd\ keep their observed value for a certain time, 
i.e. \lsf\ remains roughly constant while \lagn\ increases with time.  
In terms of the \lagn--\lsf\ diagram, it means that sources start their fastest SF growth phase at some point in the diagram 
and travel horizontally, similar to the suggestion of \cite{Shao2010}.
However, if our 5 individually detected sources were to follow this path, they would reach the AGN-dominated correlation line, 
at lower redshift, with \lagn$\geq10^{49}\,\ergs$. 
Such luminous AGN are extremely rare at $z\sim$2, and have never been observed at $z>$3 but in our sample they represent 25\%\ of the population.
The second possibility is that the five detected sources are at the peak of the SF phase and the other 20 towards its end.
Given the similar redshift of the sources, the SF must have been reduced significantly over a short period of time, while their \lagn\ remains at its peak.
The redshift range of the sample (4.65$<$z$<$4.92) implies a timescale of $\sim$100~Myr for this process.
In terms of the \lagn--\lsf\ plane, this means that the 20 sources with lower \lsf\ had higher SFR in the past but 
traveled vertically down towards the AGN-dominated correlation line. 
A possible way to rapidly quench SF is by AGN feedback.
In this scenario, which we consider more probable, feedback reduced SF in the hosts of the 20 lower-\lsf\ objects but is not yet operating efficiently in the other five.
If this scenario is true the AGN activity should be significantly reduced shortly after the quenching of the SF.
This is consistent with the estimate of the AGN duty cycle by T11.
\end{enumerate}

To conclude, our optically-selected, flux-limited sample, which consists of sources that span a narrow range of AGN properties,
provides an excellent test-bed of various SMBH and galaxy evolution scenarios at\z48.
The extremely high SFRs found in five sources provide strong evidence for a merger process in these systems.
The clear separation into two groups with SFRs that differ by a large factor provides, perhaps, indications for AGN feedback.


\begin{acknowledgements}
This work is based on observations made with Herschel, a European Space Agency Cornerstone Mission with significant participation by NASA. 
Support for this work was provided by NASA through an award issued by JPL/Caltech.
We are grateful to A. Sternberg and D. Lutz for useful discussions and comments.
An anonymous referee made useful comments that helped improve this
{\it Letter}.
We thank the DFG for support via German Israeli Cooperation grant STE1869/1-1.GE625/15-1. 
Funding for this work has also been provided by the Israel Science Foundation grant 364/07.
\end{acknowledgements}


\bibliographystyle{apj}

\end{document}